\documentclass[fleqn,usenatbib]{mnras}

\usepackage{newtxtext,newtxmath}
\usepackage[T1]{fontenc}

\DeclareRobustCommand{\VAN}[3]{#2}
\let\VANthebibliography\thebibliography
\def\thebibliography{\DeclareRobustCommand{\VAN}[3]{##3}\VANthebibliography}

\usepackage{graphicx}	% Including figure files
\usepackage{amsmath}	% Advanced maths commands
\usepackage{float}

\usepackage[dvipsnames]{xcolor}

\title[CO SLEDs in DLAs]{CO Excitation and Line Energy Distributions in Gas-selected Galaxies}

\author[A.~Klitsch et al.]{
A.~Klitsch,$^{1}$\thanks{E-mail: anne.klitsch@gmail.com}
L.~Christensen,$^{2,3}$
F.~Valentino,$^{2,3}$
N.~Kanekar,$^{4}$
P.~M{\o}ller,$^{3,5}$
M.~A.~Zwaan,$^{5}$
J.~P.~U.~Fynbo,$^{2,3}$
\newauthor{
M.~Neeleman,$^{6}$
J.~X.~Prochaska$^{7, 8}$}
\\
$^{1}$DARK, Niels Bohr Institute, University of Copenhagen, Jagtvej 128, 2200 Copenhagen N, Denmark\\
$^{2}$Cosmic Dawn Center (DAWN), Denmark,\\
$^{3}$Niels Bohr Institute, University of Copenhagen, Jagtvej 128, 2200 Copenhagen N, Denmark\\
$^{4}$National Centre for Radio Astrophysics, TIFR, Post Bag 3, Ganeshkhind, Pune 411 007, India\\
$^{5}$European Southern Observatory, Karl-Schwarzschildstrasse 2, D-85748 Garching bei M\"unchen, Germany\\
$^{6}$Max-Planck-Institut f\"ur Astronomie, K\"onigstuhl 17, D-69117, Heidelberg, Germany\\
$^{7}$Department of Astronomy \& Astrophysics, UCO/Lick Observatory, University of California, 1156 High Street, Santa Cruz, CA 95064, USA\\
$^{8}$Kavli Institute for the Physics and Mathematics of the Universe (Kavli IPMU), The University of Tokyo, 5-1-5 Kashiwanoha, Kashiwa, 277-8583, Japan
}

\date{Accepted XXX. Received YYY; in original form ZZZ}

\pubyear{2021}

\begin{document}
\label{firstpage}
\pagerange{\pageref{firstpage}--\pageref{lastpage}}
\maketitle

\begin{abstract}
While emission-selected galaxy surveys are biased towards the most luminous part of the galaxy population, absorption selection is a potentially unbiased galaxy selection technique with respect to luminosity. However, the physical properties of absorption-selected galaxies are not well characterised.
Here we study the excitation conditions in the interstellar medium (ISM) in damped Ly$\alpha$ (DLA) absorption-selected galaxies. 
We present a study of the CO spectral line energy distribution (SLED) in four high-metallicity absorption-selected galaxies with previously reported CO detections at intermediate ($z \sim 0.7$) and high ($z \sim 2$) redshifts.
We find further evidence for a wide variety of ISM conditions in these galaxies.
Two out of the four galaxies show CO SLEDs consistent with that of the Milky Way inner disk. 
Interestingly, one of these galaxies is at $z \sim 2$ and has a CO SLED below that of main-sequence galaxies at similar redshifts.
The other two galaxies at $z>2$ show more excited ISM conditions, with one of them showing thermal excitation of the mid-$J$ (J$=3, 4$) levels, similar to that seen in two massive main-sequence galaxies at these redshifts. Overall, we find that absorption selection traces a diverse population of galaxies.

\end{abstract}


%%%%%%%%%%%%%%%%%%%%%%%%%%%%%%%%%%%%%%%%%%%%%%%%%%

%%%%%%%%%%%%%%%%% BODY OF PAPER %%%%%%%%%%%%%%%%%%

\section{Introduction}

Conventional galaxy surveys identify galaxies based on the stellar or dust continuum emission and are therefore biased towards the luminous part of the galaxy population. 
Depending on the survey characteristics the emission selection tends to pick out high stellar mass galaxies with high star-formation rates (SFRs). 
An alternative and potentially unbiased galaxy selection technique with respect to stellar mass is from the absorption signature that the gas content of a galaxy and its circumgalactic medium imprints in the spectra of otherwise unrelated background quasars \citep[damped Ly$\alpha$ absorbers with an atomic hydrogen column density of N(\ion{H}{i}) $\geq2\times 10^{20}$~cm$^{-2}$;][]{Wolfe2005}. 
A good understanding of the underlying galaxy population traced by absorption selection is required to draw firm conclusions on galaxy evolution. 

Absorption-selected galaxies at intermediate redshifts ($z \sim 0.5$) were found to be extremely molecular gas-rich with low star formation efficiencies compared to emission-selected galaxies at low- ($z \sim 0$) and high ($z \sim 2$) redshifts \citep{Moller2018,Kanekar2018,Peroux2019,Szakacs2021}. 
In these studies each individual \mbox{(sub-)DLA} system was detected in a single CO rotational transition [CO(1--0), (2--1), (3--2) or (4--3)]. 
Most authors have used CO line ratios and CO-to-H$_2$ conversion factors ($\alpha_{\rm CO}$) appropriate for the Milky Way to derive molecular gas masses \citep{Kanekar2018,Moller2018,Fynbo2018,Peroux2019}. 
\citet{Szakacs2021} have used a CO line ratio appropriate for the Milky Way but derived an $\alpha_{\rm CO}$ conversion factor based on the gas metallicity.

At higher redshifts, $z \approx 2-4$, there have recently been a few detections of CO emission in absorption-selected galaxies associated with high-metallicity absorbers \citep{Neeleman2018,Neeleman2020,Fynbo2018,Kanekar2020}.
Again, each individual \mbox{DLA} system was detected in a single CO rotational transition [CO(2--1), (3--2) or (4--3)]. 
Here the authors have used a CO line ratio appropriate for galaxies near the main sequence at $z = 0 - 3$  and $\alpha_{\rm CO} =  4.36\; {\rm M_{\odot}(K \; km \; s^{-1} \; pc^2)^{-1}}$, applicable for galaxies with a near solar metallicity that are not undergoing a star burst \citep{Tacconi2020, Bolatto2013}.

\citet{Klitsch2019} studied the CO spectral line energy distribution (CO SLED) of three absorption-selected galaxies at $z \sim 0.5$. 
In two out of three galaxies they found higher excitation than in the interstellar medium (ISM) of the Milky Way. 
These galaxies have indeed been suggested to host an active nucleus and/or a starburst \mbox{\citep{Burbidge1996, Chen2005a}}, consistent with the high CO excitation.
This suggests that using line ratios from the Milky Way and $\alpha_{\rm CO}$ applicable for normal star-forming galaxies might not be appropriate for all absorption-selected galaxies. 
It is clear that the gas mass measurements critically depend on the assumptions for excitation conditions and CO-to-H$_2$ conversion factor. 
CO SLED studies are key to determine the CO excitation and thus obtain accurate estimates of the molecular gas mass.

CO SLED studies conducted so far include the Milky Way and nearby galaxies \citep[e.g.][]{Fixsen1999, Bayet2004, vanderWerf2010, Greve2014, Kamenetzky2014, Kamenetzky2017, Rosenberg2015, Lu2017}, normal star-forming galaxies at higher redshifts $z > 1.5$ \citep[][]{Daddi2015, Brisbin2019, Henriquez-Brocal2021}, lensed main-sequence galaxies \citep{Dessauges2015} and more extreme objects such as submillimetre galaxies, starburst galaxies and quasars \citep[e.g.][]{Weiss2005, Ivison2011, Bothwell2013, Harrington2021}.
It has been found that the CO excitation in high redshift galaxies is typically higher than in the inner disk of the Milky Way. 
Recently, \citet{Valentino2020a} reported CO emission line ratios for galaxies on and above the main sequence for star-forming galaxies at $z\sim 1.5$. 
They obtained CO excitation ratios above that typically observed in the inner disk of the Milky Way.
They find that the CO excitation is mainly driven by enhanced SFRs and compact galaxy sizes resulting in enhanced dense molecular gas fractions and higher dust and gas temperatures. 
Similarly, \citet{Boogaard2020} conducted a study of CO SLEDs in galaxies detected in the untargeted ALMA Spectroscopic Survey in the Hubble Ultra Deep Field \citep[ASPECS;][]{Aravena2019}. 
The galaxies in their sample lie on, above or below the main-sequence \citep{Boogaard2020, Aravena2019, Aravena2020}.
\citet{Boogaard2020} report an average CO SLED at $z = 1.2$  lower than that of the $BzK$-colour-selected galaxies at $z = 1.5$ \citep{Daddi2015} but above that of the Milky Way inner disk. 
At  higher redshift ($z = 2.5$) \citet{Boogaard2020} find a steeper CO SLED similar to that of local starburst galaxies and the low end of the excitation range observed in sub-mm galaxies (SMGs) at a similar redshift. 
According to the authors this result seems not to be due to the CO flux-selection. 
The above results suggest an intrinsic evolution of the ISM excitation conditions, with typical high-redshift galaxies having higher CO excitation than that seen in the inner disk of the Milky Way.

Absorption selection preferentially identifies lower-mass galaxies that lie on or below the main sequence of star-forming galaxies at their respective redshifts \citep[e.g.][]{Krogager2017, Kanekar2018, Rhodin2018, Rhodin2021}. 
However, the detections of CO emission in absorption-selected galaxies (especially at $z \gtrsim 2$) were obtained in the highest-metallicity absorbers, and thus may contain a bias towards object with higher stellar masses \citep{Kanekar2020}. 
It is therefore not clear if CO SLEDs inferred from emission-selected galaxy samples are appropriate for absorption-selected galaxies, too.

The goal of this study is to characterise the CO SLEDs of absorption-selected galaxies that have been detected in CO emission in previous studies, using higher-order CO rotational transitions that are observable with the Atacama Large Millimeter/submillimeter Array (ALMA).
The targets are selected irrespective of their redshift and the full sample includes six galaxies, three at intermediate redshift ($z\sim 0.5$) and three at high redshift ($z \sim 2$). 
In this work, we report observations of four galaxies, one at $z \sim 0.7$ and three at $z \sim 2 - 2.5$. 

The structure of this paper is as follows: We present the observations and data reduction in Section~2 and the source detections, CO emission line measurements, and CO line ratios in Section~3. In Section~4, we discuss the implications of our results for studies of absorption-selected galaxies.

Throughout the paper, we adopt a flat $\Lambda$-Cold Dark Matter cosmological model with $H_0 = 70$ km~s$^{-1}$~Mpc$^{-1}$, $\Omega_{\rm m} = 0.3$, and $\Omega_{\Lambda}= 0.7$ and a \citet{chabrier03} initial mass function (IMF).

\section{Observations and Data reduction} \label{Section2}

\begin{table}
	\centering
	\caption{Details of the new ALMA observations. }
	\label{tab:ObsProp}
	\begin{tabular}{lcccccccccc} % four columns, alignment for each
		\hline
DLA galaxy & CO & $\nu$ & RMS & Beam$^a$ & Beam$^b$ \\
& line & [GHz] & [mJy beam$^{-1}$] & [\arcsec~$\times$~\arcsec] & [kpc~$\times$~kpc]\\
\hline
J2335+1501 & 3--2 & 205.86 & 0.56 & 1.8 $\times$ 1.5 & 13 $\times$ 11\\
B0551--366 & 5--4 & 194.58 & 0.58 & 1.8 $\times$ 1.4 & 15 $\times$ 12\\
B0551--366 & 6--5 & 233.44 & 0.44 & 1.4 $\times$ 1.2 & 12 $\times$ 10\\
B0551--366 & 7--6 & 272.32 & 0.36 & 1.4 $\times$ 1.1 & 11 $\times$ 9\\
B1228-113 & 6--5 & 216.54 & 0.27 & 1.6 $\times$ 1.2 & 13 $\times$ 10\\
J0918+1636 & 4--3 & 128.68 & 0.47 & 1.4 $\times$ 1.1 & 11 $\times$ 9\\
J0918+1636 & 5--4 & 160.84 & 0.54 & 1.1 $\times$ 0.76 & 9 $\times$ 6\\
		\hline
	\end{tabular}
	\flushleft{
	\textit{Notes:} $\nu$ is the central observing frequency of the spectral window covering the redshifted CO line, in GHz, RMS is the RMS noise in the centre of the field quoted in mJy beam$^{-1}$ per 40~km~s$^{-1}$ channel, \mbox{$^a$}The FWHM of the clean beam, equivalent to the angular resolution in arcsec. $^b$The spatial resolution at the absorber redshift.}
\end{table}

\begin{figure*}
\includegraphics[scale = 0.5]{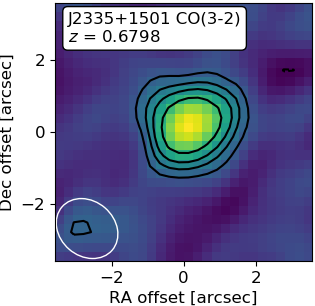} %\hspace{0.5cm}
\includegraphics[width = 0.19\linewidth]{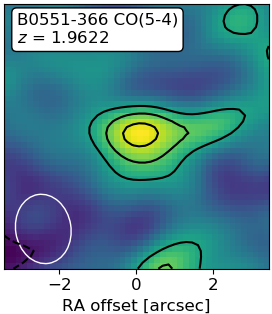}%\hspace{0.5cm}
\includegraphics[width = 0.19\linewidth]{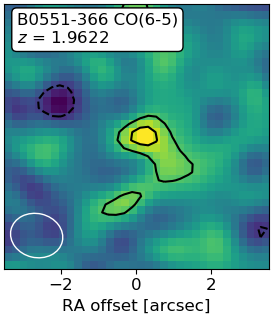}
\includegraphics[width = 0.19\linewidth]{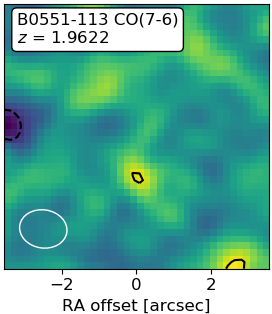}\\%\vspace{0.5cm}
\includegraphics[scale = 0.5]{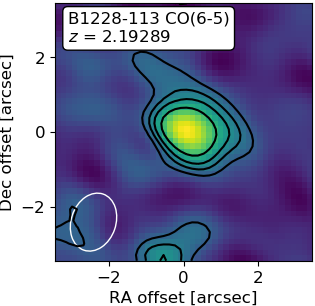}%\hspace{0.5cm}
\includegraphics[width = 0.19\linewidth]{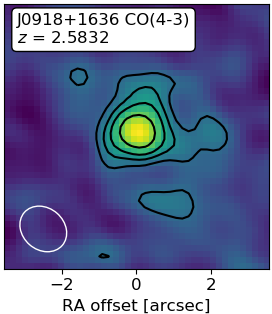}
\includegraphics[width = 0.19\linewidth]{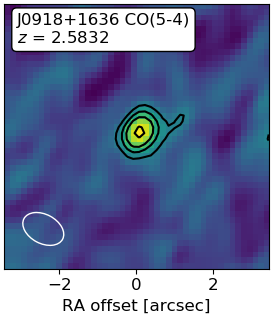}
\caption{Integrated CO line flux density maps of the four detected galaxies. The maps are calculated by integrating over the velocity channels marked in yellow in Fig.~\ref{Fig:Spectra}. For the non-detected CO(7--6) transition in DLA~B0551-113 we integrate over the same velocity range as for the CO(6--5) transition. Black contours mark the 3, 5, 7 and 10~$\sigma$ levels. Negative contours are shown as dashed lines. The white ellipse shows the synthesized beam. \label{Fig:Mom0Maps}}
\end{figure*}

\begin{figure*}
\includegraphics[scale = 0.5]{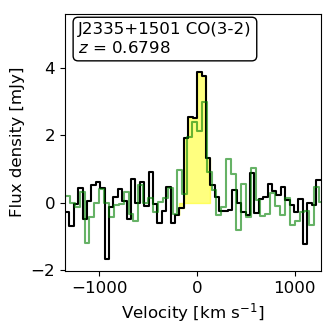}
\includegraphics[scale = 0.5]{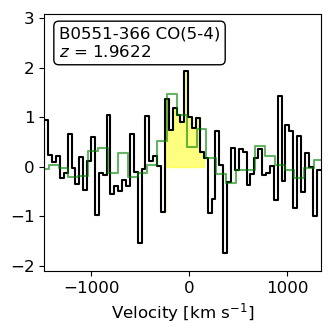}
\includegraphics[scale = 0.5]{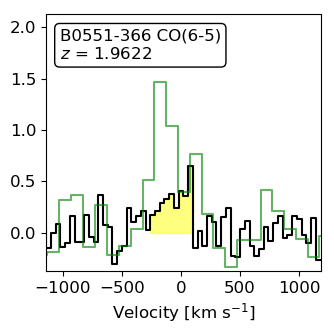}
\includegraphics[scale = 0.5]{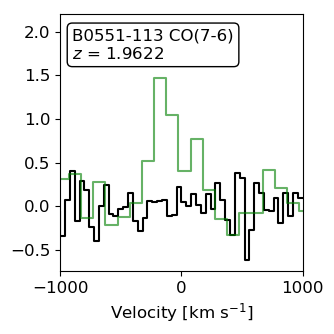}\\
\includegraphics[scale = 0.5]{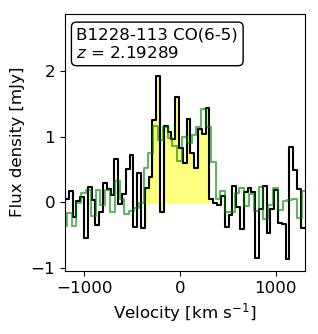}
\includegraphics[scale = 0.5]{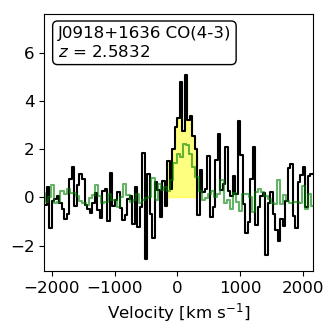}
\includegraphics[scale = 0.5]{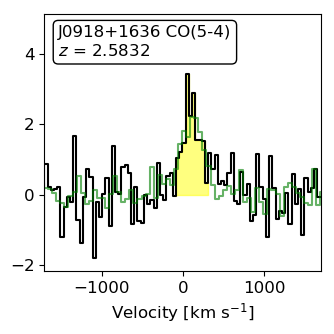}
\caption{CO emission line spectra of the four detected galaxies. As a comparison we show the new CO line observations from this work (black) and the previously observed CO lines \citep[green; ][]{Kanekar2018, Kanekar2020}. The yellow area marks the channels used for calculating the integrated flux density. It can be seen that the line centres and widths measured from the different CO transitions agree well. For CO(7--6) from B0551--113 we show the spectrum extracted within one beam at the position of the absorber host galaxy. \label{Fig:Spectra}}
\end{figure*}

\begin{figure}
\includegraphics[width = \linewidth]{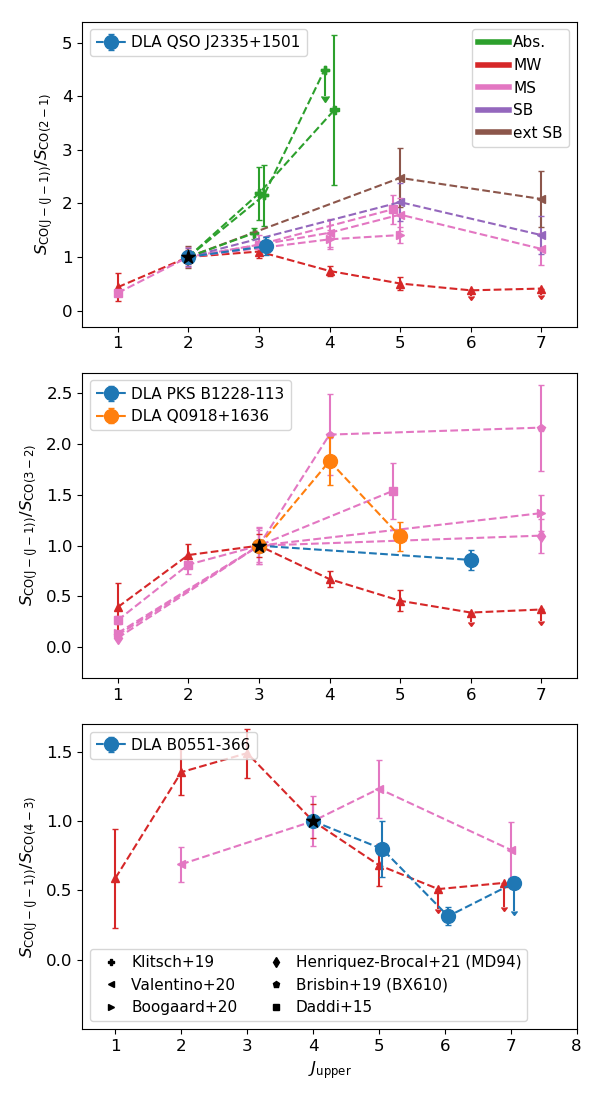}
\caption{CO SLEDs of absorption-selected galaxies. 
We note that due to the missing CO(1--0) flux density measurement these plots are normalized to the lowest available CO transition (black star). 
The CO SLEDs of the inner disk of the Milky Way \citep[red; ][]{Fixsen1999}, other absorption-selected galaxies at $z\sim 0.5$ \citep[green; ][]{Klitsch2019}, main-sequence galaxies at $z = 1.2 - 2.2$ \citep[pink; ][]{Daddi2015,Brisbin2019,Boogaard2020,Valentino2020a,Henriquez-Brocal2021}, starburst (purple) and extreme starburst (brown) galaxies at $z \sim 1.5$ \citep{Valentino2020a} are shown for comparison (colours differentiate object classes, while symbols indicate literature references). 
Arrows show the three sigma upper limits.
We find that absorption-selected galaxies cover a broad range of excitation conditions. \label{Fig:COSLED}}
\end{figure}

We observe one or more higher $J$ rotational transitions from a sample of absorption-selected galaxies previously detected in CO emission \citep{Neeleman2018, Fynbo2018, Kanekar2018, Kanekar2020} to study the excitation conditions in the ISM using the CO SLEDs. 
The original targets were selected from damped or sub-damped Ly$\alpha$ absorption systems
%with N(HI)~$= 5 \times 10^{19} \ {\rm cm ^2}$ for the intermediate-redshift absorber, and N(HI)~$\geq 3 \times 10^{20} \ {\rm cm ^2}$ for the high-redshift absorber], 
having relatively high metallicities, above 10 per cent solar. 
This is because high-metallicity absorbers are expected to be associated with more massive galaxies \citep{Moller2013,Christensen2014}, which are consequently expected to have both larger molecular gas masses and lower $\alpha_{\rm CO}$ values. 
CO emission was typically detected in the fields of absorbers with the highest metallicities, [M/H]~$\gtrsim -0.25$. 
The detection fraction of CO emission in these absorption-selected galaxies is 5/7 in the intermediate-redshift sample \citep{Kanekar2018} and 5/12 in the high-redshift sample \citep{Kanekar2020}. 
Targets with non-detections of CO emission have molecular gas masses  $< 5 \times 10^{9} {\rm M_{\odot}}$ and $< 10^{10} {\rm M_{\odot}}$ in the intermediate-$z$ and high-$z$ samples, respectively \citep{Kanekar2018,Kanekar2020}.

The ALMA observations are carried out under program 2019.1.01457.S (PI: L.~Christensen) and the observed fields and transitions are listed in Table~\ref{tab:ObsProp}. 
Two fields are observed in one new CO transition, one field in two CO transitions, and one in three CO transitions. 
All observations use four 1.875 GHz wide spectral windows, with one spectral window (in FDM mode of the correlator) covering the expected redshifted CO line frequency, and the other three spectral windows (in TDM mode) placed at neighbouring frequencies to measure the continuum emission.

We use the pipeline calibrated $uv$-data sets, as delivered by the Joint ALMA Observatory. 
Additional data reduction steps are carried out with the Common Astronomy Software Applications (\textsc{CASA}) software package version 5.6 \mbox{\citep{McMullin2007}}. 
In the case of PKS~B1228--113, the QSO continuum flux density is sufficient to self-calibrate the visibilities, and thus improve our estimates of the antenna-based gains over those obtained from the pipeline calibration. 
The self-calibration is done using the routine {\sc gaincalR} \citep{Chowdhury2020}, following a standard iterative imaging and self-calibration procedure. 
For all fields, continuum images are made from the line-free spectral windows with the {\sc tclean} routine, using ``Briggs'' weighting \citep{Briggs95} with a ``robust'' parameter of 0.5. 
The continuum is subtracted from the calibrated spectral-line visibilities using the routine {\sc uvsub}. 
The continuum-subtracted visibilities are then imaged, at a velocity resolution of $40$~km~s$^{-1}$, cleaning down to $0.5\sigma$, where $\sigma$ is the RMS noise per $40$~km~s$^{-1}$ channel. 
Natural weighting is used, to obtain
%The imaging is done using the standard \textit{tclean} routine with a final field of view of 1.5 times the primary beam size. 
the maximum sensitivity, appropriate for a line detection experiment for unresolved sources. 
Any residual continuum emission is subtracted from the spectral cubes by fitting a first-order polynomial to line-free channels, using the routine {\sc imcontsub}. 
The final beam sizes as well as the RMS noise per 40~km~s$^{-1}$ channel at the centre of each field are listed in Table \ref{tab:ObsProp}. 
The systematic uncertainty on the flux density scale in the ALMA bands used here is assumed to be 5 per cent (ALMA Technical Handbook\footnote{\url{https://almascience.nrao.edu/documents-and-tools/cycle7/alma-technical-handbook}}), which we interpret as a 2$\sigma$ error.
%\textbf{We create image cubes for which we subtract the continuum emission using the task \textit{uvsub} and \textit{imcontsub} to measure the emission line flux. Additionally, we create continuum images from the line free channels.}

\begin{table*}
	\centering
	\caption{Overview of physical properties of the absorption-selected galaxies.}
	\label{tab:GalOverview}
	\begin{tabular}{lcccccccccc} % four columns, alignment for each
		\hline
		DLA galaxy & $z_{\rm QSO}$ & $z_{\rm abs}$ & RA & Dec & CO & $b$ & FWHM & $S_{\rm int}$ & $S_{\rm peak}$ & $L'_{\rm CO}\times 10 ^{9}$\\
		& & & & & line & [kpc] &  [km s$^{-1}$] & [Jy km s$^{-1}$] & [mJy] &  [K km s$^{-1}$ pc$^2$] \\
		\hline
J2335+1501 & 0.791 & 0.6798 & 23h35m44.19s & +15d01m14.3s & 3--2 & 29  & 226 $\pm$ 45 & 0.722 $\pm$ 0.066 & 3.88 $\pm$ 0.50 & 1.97 $\pm$ 0.18  \\
B0551--366 & 2.317 & 1.9622 &  05h52m46.17s & --36d37m25.9s & 5--4 & 14 & 357 $\pm$ 40 & 0.346 $\pm$ 0.078 & 1.93 $\pm$ 0.60 & 2.64 $\pm$ 0.60 \\
 & & &  &  & 6--5 & & 239 $\pm$ 40 & 0.136 $\pm$ 0.023 & 0.76 $\pm$ 0.21 & 0.72 $\pm$ 0.12  \\
 & & &  &  & 7--6 & & n.a. & $<0.24$ & n.a. & $<0.93$  \\
B1228--113 & 3.528 & 2.1929 & 12h30m57.07s & --11d39m29.0s & 6--5 & 28 & 595 $\pm$ 40 &  0.624 $\pm$ 0.064 & 1.92 $\pm$ 0.37 & 4.01 $\pm$ 0.41  \\
J0918+1636 & 3.096 & 2.5832 & 09h18m26.26s & +16d35m55.4s & 4--3 & 117 & 320 $\pm$ 40 & 1.35 $\pm$ 0.15 & 5.1 $\pm$ 1.0 & 25.9 $\pm$ 2.9 \\
% & & &  &  & 5--4 & & 198 $\pm$ 49 & 0.85 $\pm$ 0.10 & 3.03 $\pm$ 0.68 & 10.49 $\pm$ 1.26 \\
 & & &  &  & 5--4 & & 119 $\pm$ 40 & 0.803 $\pm$ 0.092 & 3.43 $\pm$ 0.69 & 9.9 $\pm$ 1.1 \\
		\hline
	\end{tabular}
	\flushleft{\textit{Notes:} $b$ denotes the impact parameter between the quasar sightline and the position of the galaxy, FWHM is the width of the CO emission line at 50 percent of the peak flux density, sometimes also referred to as $W_{50}$, $S_{\rm int}$ is the integrated line flux density, $S_{\rm peak}$ is the peak flux density of the line, $L'_{\rm CO}$ denotes the line luminosity of the observed transition, errors on the last three quantities include a 2.5 per cent error on the absolute flux density scale. We quote a 3$\sigma$ upper limit for the CO(7--6) emission line flux density in DLA~B0551-366 adopting the same FWHM as in the CO(6--5) emission line. }
\end{table*}

\section{Analysis}

We perform an automatic source finding on the spectral windows covering the targeted CO line using the \textsc{SoFiA} source finding application \citep{Serra2015} developed for \ion{H}{i} interferometric observations. 
We use the standard configuration of the `Smooth + Clip Finder' with a 4\,$\sigma$ detection threshold. 
We detect all known absorption-selected galaxies in each field in all observed CO transitions except for the CO(7--6) transition in DLA~B0551-366. 
Apart from these previously reported galaxies, we do not find any additional galaxies. 

In particular, we search for CO emission from the low-impact parameter ($b \approx 16$~kpc) absorption-selected galaxy in the J0918+1636 field \citep{Fynbo2013}.
This galaxy was was not detected in \mbox{CO(3--2)} emission by \citet{Fynbo2018}. 
However, we do not detect \mbox{CO(4--3)} emission from this galaxy.
\citet{Fynbo2018} report another galaxy at the same redshift as the absorber, located at a larger impact parameter of 117~kpc which we here refer to as DLA~J0918+1636.

We measure the CO line flux densities of the previously reported galaxies with CO detections in order to examine their CO SLEDs. 
Integrated CO line flux density maps and CO spectra of the galaxies are shown in Figs.~\ref{Fig:Mom0Maps} and \ref{Fig:Spectra}, respectively. 
We determined the optimal velocity range for the flux density maps covering the entire CO emission line in an iterative manner. 
The spectra are extracted from a region defined by the $3 \sigma$ contour in the integrated flux density images. 
Details of the detections are listed in Table~\ref{tab:GalOverview}. 
The line full-widths-at-half-maximum (FWHMs) are determined directly from the extracted spectra shown in Fig.~\ref{Fig:Spectra}, by measuring the width of the line profile at 50 per cent of the peak flux density.

We compare the spectral line shape of the different CO transitions reported in this work with those of the lower-$J$ lines in the literature (see Fig.~\ref{Fig:Spectra}). 
The line centres and FWHMs of the various transitions reported in Table~\ref{tab:Comparison} agree within the uncertainties with the measurements from the lower-$J$ lines.

\begin{table*}
	\centering
	\caption{Comparison to previous observations and CO line luminosity ratios.}
	\label{tab:Comparison}
	\begin{tabular}{lccccccccccc}
		\hline
		DLA galaxy & CO & FWHM & $S_{\rm int}$ & $S_u/S_l$ & $L'_{\rm CO}\times 10 ^{9}$ & $r_{\rm ul}$ & $r_{\rm ul}^{\rm MW}$ & Ref. \\
		& line & [km s$^{-1}$] & [Jy km s$^{-1}$] & & [K km s$^{-1}$ pc$^2$] \\
		\hline
J2335+1501 & 2--1 & 250 $\pm$ 50 & $0.601 \pm 0.064$ & 1 & $3.70 \pm 0.40$ & 1 & 1 & \citet{Kanekar2018}\\
& 3--2 & 226 $\pm$ 45 & 0.722 $\pm$ 0.066 & $1.20 \pm 0.17$ & 1.97 $\pm$ 0.18 & $0.532 \pm 0.075$ & $0.490\pm 0.058$ & This work\\
B0551--366 & 4--3 & $300 \pm 25$ & $0.433 \pm 0.052$ & 1 & $5.15 \pm 0.62$ & 1 & 1 & \citet{Kanekar2020}\\
 & 5--4 & 357 $\pm$ 40 & 0.346 $\pm$ 0.078 & $ 0.80 \pm 0.20$ & 2.64 $\pm$ 0.60 & $ 0.51 \pm 0.13 $ & $0.437 \pm 0.098$ & This work \\
 & 6--5 & 239 $\pm$ 40 & 0.136 $\pm$ 0.023 & $0.314 \pm 0.065$ & 0.72 $\pm$ 0.12 & $0.140 \pm 0.029$ & $< 0.23$ & This work \\
 & 7--6 & n.a. & $<0.24$ & $<0.55$ & $<0.93$ & $<0.17$ & $<0.18$ & This work\\
B1228--113 & 3--2 & $600 \pm 22$ & $0.726 \pm 0.031$ & 1 & $18.69 \pm 0.80$ & 1 & 1 & \citet{Neeleman2018}\\
 & 6--5 &  595 $\pm$ 40 & 0.624 $\pm$ 0.064 & $0.860 \pm 0.095$ & 4.01 $\pm$ 0.41 & $0.215 \pm 0.024$ & $< 0.086$ & This work\\
J0918+1636  & 3--2 & $350 \pm 25$  & $0.736 \pm 0.045$ & 1 & $25.1 \pm 1.6$ & 1 & 1 & \citet{Fynbo2018}\\
& 4--3 & 320 $\pm$ 40 & 1.35 $\pm$ 0.15 & $1.83 \pm 0.24$ & 25.9 $\pm$ 2.9 & $1.03 \pm 0.13$ & $0.377 \pm 0.045$ & This work\\
& 5--4 & 119 $\pm$ 40 & 0.803 $\pm$ 0.092 & $ 1.09 \pm 0.14 $ & 9.9 $\pm$ 1.1 & $ 0.394\pm 0.051 $ & $0.165 \pm 0.037$ & This work\\
		\hline
	\end{tabular}
	\flushleft
	\textit{Notes:} FWHM denotes the width of the line at 50 per cent of the peak line flux density. $S_{\rm int}$ is the integrated line flux density. $S_u/S_l$ is the line flux density ratio that is also plotted in Fig.~\ref{Fig:COSLED}. $L'_{\rm CO}$ is the line luminosity of the given CO transition; values from \cite{Kanekar2018} and \citet{Kanekar2020} are scaled to the cosmology used in this paper. $r_{\rm ul}$ is the line luminosity ratio between the CO transition in the respective row and the lowest observed CO transition for that galaxy; $r_{\rm ul}^{\rm MW}$ denotes the same line luminosity ratio measured in the inner disk of the Milky Way \citep{Fixsen1999}. Limits on the line luminosity ratio are $3\sigma$ upper limits.
\end{table*}

\subsection{CO Excitation and Line Energy Distribution}

Using the previously reported mid-$J$ rotational transitions we calculate the CO line flux density and luminosity ratios of the different available transitions for each absorber. 
The CO(1--0) transition that is traditionally used for this comparison is not available for any of our targets. 
We therefore calculate the line luminosity ratios $L'_{\text{CO}(u-[u-1])}/L'_{\text{CO}(l-[l-1])} = r_{ul}$, where $u$ is the upper $J$ level of the higher available rotational transition and $l$ is the upper $J$ level of the lower available transition. 
The line ratios are reported in Table~\ref{tab:Comparison}, where the reported uncertainties on the line ratios include the flux density errors in each transition.
The same ratios for the inner disk of the Milky Way based on CO line observations \citep{Fixsen1999} are also listed in Table~\ref{tab:Comparison}. 
For high $J$ CO transitions [CO(6--5) and CO(7--6)] in the Milky Way inner disk we list the 3$\sigma$ upper limits \citep{Fixsen1999}.

The line ratios in two out of our four absorption-selected galaxies (DLA~J2335+1501 and DLA~B0551--366) are consistent with the same line ratios for the Milky Way inner disk. 
The line luminosity ratios in the absorption-selected galaxies DLA~B1228--113 and DLA~J0918+1636 are significantly above that of the Milky Way inner disk.
%DLA~J0918+1636 has $r_{43}$ significantly above that of the Milky Way inner disk while $r_{53}$ is consistent with $r_{53}^{\rm MW}$.}
However, we note that line ratios involving neighbouring $J$ transitions can be ambiguous especially in the mid- and high-$J$ transitions where the gradient in the excitation ladder is smoother than at low $J$.

In addition to the Milky Way CO line luminosity ratios, we compare the line flux density ratios in Table~\ref{tab:Comparison} with the CO SLEDs of different galaxy types in Fig~\ref{Fig:COSLED}. 
The CO SLEDs of absorption-selected galaxies from this work are compared to the CO SLEDs of three absorption-selected galaxies at $z \sim 0.5$ \citep{Klitsch2018,Klitsch2019}, the inner disk of the Milky Way \citep{Fixsen1999}, main-sequence galaxies at $z = 1.2-2.2$ \citep{Daddi2015, Brisbin2019, Boogaard2020, Valentino2020a, Henriquez-Brocal2021}, starburst and extreme starburst galaxies at $z \sim 1.5$ \citep{Valentino2020a}. 
Again, we anchor the CO SLED to the lowest available transition for the different absorption-selected galaxies. 

It can be seen that two out of the four CO SLEDs of absorption-selected galaxies have low excitation conditions consistent with that of the Milky Way inner disk and the other two show higher excitation. 
In the following we analyse the CO SLEDs of the different galaxies in more detail. 

The lowest redshift galaxy in our sample, DLA~J2335+150 at $z=0.6798$, has a ratio that is consistent within the errors with the CO SLED of the Milky Way inner disk, but is also consistent with more excited CO SLEDs reported for galaxies on and above the star-forming main sequence at $z \sim 1.2 - 1.5$ \citep{Daddi2015,Boogaard2020,Valentino2020a}.
Measurements of higher $J$ transitions are required for this galaxy to test whether the CO SLED is similar to that of the Milky Way inner disk.
The CO SLED of DLA~B0551--366 at $z = 1.9622$  is consistent with the $3\sigma$ upper limits of the Milky Way inner disk but below the average CO SLED of main sequence galaxies at $z\sim 1.5$  \citep{Valentino2020a}.
The CO SLED of DLA~B1228--113 at $z=2.1929$ is below the CO SLED of main sequence galaxies at $z \sim 1.5$ and $z \sim 2.5$ but above that of the Milky Way inner disk. Finally, the CO SLED of DLA~J0918+1636 at $z = 2.5832$ has an approximately thermal $r_{43}$ while $r_{53}$ is less than unity, but significantly above $r_{53}^{\rm MW}$. This suggests that the CO SLED of  DLA~J0918+1636 is likely to be thermalised up to the $J=4$ rotational level.
We note that these are neighbouring mid-$J$ CO transitions and that the main-sequence galaxy BX610 shows a similar $r_{43}$ luminosity ratio, also consistent with thermal excitation \mbox{\citep{Brisbin2019}. }
Further, observations of the lower-$J$ lines are not available for DLA~B0551--366 and of the intermediate-$J$ lines for DLA~B1228--113. We are therefore unable to determine the slope and the peak of the CO SLEDs in these galaxies.
A better sampling of the CO SLED in these galaxies, as well as observations of the CO~$J(1-0)$ line in all our targets, are crucial to better understand the CO line excitation.

CO SLEDs are non-linear, non-monotonic functions with changing slopes as a function of the upper-state rotational quantum number $J$ and different maxima for different object classes. 
Therefore we caution that differences in the CO SLEDs as well as line luminosity ratios $r_{ul}$ anchored on higher $J$ rotational transitions (above the CO(1--0) transition) can appear less pronounced. 
As an example $r_{74}$ of the main-sequence galaxies at $z \sim 1.2$ \citep{Valentino2020a} is consistent with that of the Milky Way inner disk while the $r_{72}$ ratio is above that of the Milky Way inner disk. 
Additionally, the incomplete sampling of CO SLEDs in different types of galaxies does not allow a direct comparison between all object classes. 
Even the class of main-sequence galaxies presented in Fig.~\ref{Fig:COSLED} (middle panel) shows a wide range of excitation conditions.

\subsection{SED Fitting}

In addition to the CO line emission, we detected continuum emission from the absorber host galaxies DLA~B1228--113 and DLA~J0918+1636. 
The continuum images are shown in Fig.~\ref{FigContImage} and the flux density measurements are listed in Table~\ref{tab:appendix_phot}.

We used {\tt MAGPHYS}, \citep[optimised for high-redshift galaxies;][]{daCunha2008, daCunha2015} to carry out a fit to the broad-band spectral energy distributions (SEDs) of DLA~J0918+1636 and DLA~B1228--113, assuming a \citet{chabrier03} IMF.
All input values for the SED fits are listed in Table~\ref{tab:appendix_phot}, the SED fit is shown in Fig.~\ref{Fig:appendixfig} and the best fitting results are listed in Table~\ref{tab:magphys_out}.

For DLA~J0918+1636 at $z = 2.5832$ galaxy, we combine the new ALMA 2mm flux density with the existing optical and near-IR photometry \citep{Fynbo2018}. 
%and use {\tt MAGPHYS} \citep[optimised for high-redshift galaxies;][]{daCunha2008, daCunha2015} to carry out a fit to the spectral energy distribution (SED).  
%Assuming a \citet{chabrier03} IMF. 
We obtain a stellar mass of log$(M_{\star}/M_{\odot}) = 10.49^{+0.16}_{-0.11}$, consistent with the estimate in \mbox{\citet{Fynbo2018}.} 
Our fit yields an SFR of $229^{+151}_{-91} \ {\rm M}_\odot$~yr$^{-1}$, higher than the earlier estimate, albeit consistent within the errors.
%%%%%%%%%%%%%%%%%%%%%%%%%%%%%%%%%%%%%%%%%%%%%%%%%%%%%%%%%%%%%%%%%%%%%%%%
% -> Does this include the IR emission? In this case, it makes sense, if Johan just used the UV. % 
%  LC :  I used the same data as in Fynbos paper, but with one additional ALMA data point.
%%%%%%%%%%%%%%%%%%%%%%%%%%%%%%%%%%%%%%%%%%%%%%%%%%%%%%%%%%%%%%%%%%%%%%%%

For DLA~B1228--113 at $z = 2.1929$, we combine our ALMA 1.3~mm flux density measurement with the 3~mm measurement of \citet{Neeleman2018}, and use Very Large Telescope Multi Unit Spectroscopic Explorer (VLT/MUSE) archival data (programme ID 197.A-84, PI: Fumagalli) for the optical photometry. From the MUSE data cube, we extract a one-dimensional spectrum at the location of the galaxy within a 1\arcsec~aperture diameter and integrate this with the Sloan Digital Sky Survey (SDSS) filter transmission curves to derive AB magnitudes $g=25.3\pm0.3$, $r=24.7\pm0.2$ and $i=24.4\pm0.3$ for the galaxy. The photometry is corrected for a Galactic reddening  $E_{B-V}=0.0335$ \citep{Schlafly2011}.
%\textbf{All input values  for the SED fitting are listed in Table~\ref{tab:appendix_phot}, the SED fit is shown in Fig.~\ref{Fig:appendixfig} and the best fitting results are listed in Table~\ref{tab:magphys_out}.}
The best SED fit with {\tt MAGPHYS} yields a stellar mass of $\log(M_*/M_{\odot}) = 10.13^{+0.34}_{-0.23}$ and a total SFR of $87^{+39}_{-28} \; {\rm M_{\odot} \; yr^{-1}}$, in agreement with the dust-corrected SFR derived from the H$\alpha$ emission \citep{Neeleman2018}.

%%%%%%%%%%%%%%%%%%%%%%%%%%%%%%%%%%%%%%%%%%%%%%%%%%%%%%%%%%%%%%%%%%%%%%%%%%%
% Here I assume that we'll keep Magphys for reference, but trust the simplified MBB for comparison %
% with Scoville+, Kaasinen+.                                                                                                                 %
%%%%%%%%%%%%%%%%%%%%%%%%%%%%%%%%%%%%%%%%%%%%%%%%%%%%%%%%%%%%%%%%%%%%%%%%%%%

{\tt MAGPHYS} also returns dust masses of $\log(M_{\rm dust}/$M$_{\odot}) = 8.74^{+0.10}_{-0.12}$ for DLA~B1228--113, and $\log(M_{\rm dust}/$M$_{\odot})= 8.62^{+0.16}_{-0.15}$ for DLA~J0918+1636. 
These are inferred from the optical-to-sub-mm photometry by assuming energy balance. However, we currently do not have any measurements probing the peak of the far IR emission and, thus, the dust temperature and total IR luminosity are not well-constrained.

To avoid possible overfitting of the far-IR regime, we derived an alternative estimate of $M_{\rm dust}$ by simply rescaling the observed photometry in the mm regime --- probing the Rayleigh-Jeans tail of the dust emission --- with a modified black body curve \mbox{(MBB; \citealt{Berta2016})}. 
We assumed $\beta=1.8$ \mbox{\citep[e.g.][]{Berta2016, Scoville2016}} and a dust mass absorption coefficient of $k_{0} = 5.1\,\mathrm{cm^2\,g^{-1}}$ at rest-frame $\nu_0=250\; \mu$m  consistent with \citet{Magdis2012a}. 
For $T_{\rm dust}=35$ K typical for galaxies at $z\sim2-2.5$ (\citealt{Schreiber2018} and references therein), we find $\log(M_{\rm dust}/$M$_{\odot}) = 8.83 \pm 0.28$ for DLA~B1228--113 and $\log(M_{\rm dust}/$M$_{\odot})= 8.72 \pm 0.27 $ for DLA~J0918+1636. 
The uncertainties mark the maximal fluctuation of $M_{\rm dust}$ obtained by allowing for $T_{\rm dust}=35\pm5$ K and $\beta=1.8\pm0.2$. Variations in the dust mass absorption coefficient could induce an extra uncertainty of $\gtrsim 0.18$~dex \citep{Andersen2007}. 
Statistical uncertainties on the photometric measurements are $\approx 0.09$~dex and $\approx 0.14$~dex for DLA~B1228--113 and DLA~J0918+1636, respectively.

The MBB rescaling also allows us to estimate the luminosity at rest-frame 850-$\mu$m, $L_{\rm 850\mu m}$, recently proposed as a dust (and gas) mass tracer \mbox{\citep{Scoville2016}}. Here we use $L_{\rm 850\mu m}$ to predict the expected $L'_{\rm CO(1-0)}$ luminosity, assuming that our targets follow the same relation within a factor of two \citep{Kaasinen2019}. This yields $\log(L_{\rm 850\mu m}/\mathrm{erg\,s^{-1}\,Hz^{-1}})=30.80 \pm 0.12$ and $30.88\pm0.13$ for DLA~B1228--113 and DLA~J0918+1636, respectively, assuming the same ranges of $T_{\rm dust}$ and $\beta$ as above.

\subsection{Dust-Based CO Luminosities and Molecular Gas Masses}

\begin{table*}
	\centering
	\caption{Molecular gas masses from different methods.}
	\label{tab:MolGasMass}
\begin{tabular}{lcccccccc}
\hline
DLA galaxy & $L_{\rm CO(1-0)}^{' \; \text{mid-}J \; \text{CO}}  \times 10^{10}$ & $L_{\rm CO(1-0)}^{' \; \text{mid-}J \; \text{CO}} \times 10^{10}$ &$L_{\rm CO(1-0)}^{\rm ' \;850 \mu m}\times 10^{10}$ & $M_{\rm mol}^{\rm 850 \mu m} \times 10^{10}$ &  $M_{\rm mol}^{\rm CO}\times 10^{10}$ &  $M_{\rm mol}^{\rm CO}\times 10^{10}$ & $M_{\rm mol}^{\rm dust} \times 10^{10}$ \\
& $(r_{31} = 0.55)$ & $(r_{31} = 1)$ & & & $(r_{31} = 0.55)$ & $(r_{31} = 1)$\\
& [K km s$^{-1}$ pc $^{2}$] & [K km s$^{-1}$ pc $^{2}$] & [K km s$^{-1}$ pc $^{2}$] & [$ \text{M}_{\odot}$] & [$ \text{M}_{\odot}$] & [$ \text{M}_{\odot}$]& [$ \text{M}_{\odot}$]\\
\hline
B1228-113 &  $3.40 \pm 0.15$ & $1.869 \pm 0.080$ & $1.91^{+0.78}_{-0.67}$ & $8.3 ^{+3.4}_{-2.9}$ & $14.82 \pm 0.65$ & $8.15 \pm 0.35$ & $4.6^{+8.5}_{-3.5}$ \\
J0918+1636 & $4.56 \pm 0.29$ & $2.51 \pm 0.16$ & $2.29^{+0.99}_{-0.83}$ &$9.98^{+4.3}_{-3.6}$ & $19.9 \pm 1.3$ & $10.94 \pm 0.70$ & $2.8^{+5.6}_{-1.8}$  \\

\hline
\end{tabular}
\flushleft
	\textit{Notes:} $L_{\rm CO(1-0)}^{\rm ' \;850 \mu m}$ is the CO(1--0) luminosity derived from the rest frame 850$\mu$m luminosity using the $L_{\rm CO(1-0)}^{\rm ' \;850 \mu m}$ - $L_{\rm \;850 \mu m}$ scaling relation of \citet{Scoville2016}, $L_{\rm CO(1-0)}^{' \; \text{mid-}J \; \text{CO}}$ is the CO(1--0) luminosity derived from the mid-$J$ transitions \citep{Kanekar2020}. We provide estimates of $L_{\rm CO(1-0)}^{' \; \text{mid-}J \; \text{CO}}$ and the molecular gas mass $M_{\rm mol}^{\rm CO}$ based on two different values of the CO line luminosity ratio $r_{31}$, assuming $r_{31}= 0.55, 1.0$. $M_{\rm mol}^{\rm dust}$ is derived from an assumed dust-to-gas ratio following \citet{Magdis2012a}. 
%$M_{\rm mol}^{\rm CO}$ is derived from the CO emission lines reported by \citet{Kanekar2020} \textbf{also recalculated using the same cosmological model}. 
See text for a discussion of these values. All quantities have been scaled to the cosmology assumed in this paper.
\end{table*}

We derive the CO(1--0) luminosity, $L_{\rm CO(1-0)}^{\rm '}$, based on the rest frame 850$\mu$m luminosity following the prescription of \citet{Scoville2016}. 
% We refrain from converting to molecular gas masses to avoid additional complications introduced by a choice of $\alpha_{\rm CO}$ and focus instead on a comparison of $L_{\rm CO(1-0)}^{\rm ' \;850 \mu m}$ predicted from the dust emission and $L_{\rm CO(1-0)}^{' \; \text{mid-}J \; \text{CO}}$ derived from scaling mid-$J$ CO transitions and assuming CO excitation conditions. 
This allows us to compare the $L_{\rm CO(1-0)}^{\rm '}$ value inferred from the dust continuum emission (i.e. $L_{\rm CO(1-0)}^{\rm ' \;850 \mu m}$ with that ($L_{\rm CO(1-0)}^{' \; \text{mid-}J \; \text{CO}}$) derived from the measured luminosity of the mid-$J$ CO transitions.
The results are presented in Table~\ref{tab:MolGasMass}. We note that 
\citet{Scoville2016} argue that a single scaling factor is applicable for galaxies with a stellar mass of $M_{\star} > 2 \times 10^{10} \text{M}_{\odot}$ irrespective of redshift and galaxy type.  

%We find that the CO luminosity for DLA~PKS~B1228-113 derived from the two different methods are consistent. 
%\citet{Kanekar2020} used CO line ratios appropriate for average main-sequence galaxies at that redshift. 
%We show in Fig.~\ref{Fig:COSLED} and Table~\ref{tab:Comparison} that DLA~PKS~B1228-113 has, indeed, excitation conditions similar to main-sequence galaxies at that redshift.

For both galaxies, the CO luminosities derived from the two methods do not agree within the errors. 
While \citet{Kanekar2020} used sub-thermal CO line ratios to infer the CO(1--0) line luminosity, we show in Fig.~\ref{Fig:COSLED} and Table~\ref{tab:Comparison} that DLA~J0918+1636 shows thermalised CO excitation conditions for the $J = 4$ level relative to the $J = 3$ level. 
This suggests that, both mid-$J$ levels would also show thermal excitation relative to the $J=1$ level (i.e. that $r_{41}\approx 1$ and $r_{31} \approx 1$, rather than $r_{31} \approx 0.55$, as assumed by \citealt{Kanekar2020}).
Adopting this line ratio results in a factor of $\approx 1.8$ lower $L'_{\rm CO(1-0)}$ than the estimate of \citet{Kanekar2020}, i.e. $L'_{\rm CO(1-0)} \approx (2.51 \pm 0.16) \times 10^{10}$~K~km~s$^{-1}$~pc$^2$. 
This is consistent with the estimate of the CO(1--0) line luminosity from the rest frame 850-$\mu$m luminosity.
For DLA~B1228-113 our observations only cover the CO(3--2) and CO(6--5) transitions. However, the agreement between the $L_{\rm CO(1-0)}^{\rm ' \;850 \mu m}$ and $L_{\rm CO(1-0)}^{' \; \text{mid-}J \; \text{CO}}$ (assuming $r_{31} = 1$) suggests that the excitation of the low-$J$ CO rotational levels is likely to be near thermal.

Next, we follow the description in \citet{Magdis2012a} to derive the metallicity-dependent dust-to-gas mass ratio $\delta_{\rm GDR}$. 
We note that the absorption metallicities of the two DLAs are known \citep{moller2020,Kanekar2020}, but these correspond to the pencil beam along the QSO sightline. Due to metallicity gradients \citep[e.g.][]{Christensen2014}, the metallicity of the host galaxy in which the molecular gas is located may be different from that measured in absorption.
For consistency with the approach in \citet{Magdis2012a}, we estimate the metallicities from the Fundamental Mass Metallicity relation \citep{Mannucci2010} rescaled to the \citet{Pettini2004} metallicity scale. 
We propagate the uncertainties on the best-fit $M_{\star}$ and SFR values, and the $\sim$0.2~dex uncertainty on the metallicity calibration to $\delta_{\rm GDR}$.  
We derive $12+ \log \mathrm{(O/H)} = 8.40 \pm 0.2$ and $8.49 \pm 0.2$ ($[\text{M/H}]=-0.29$
and $-0.20$, respectively) for DLA~B1228--113 and DLA~J0918+1636, respectively.
\footnote{For reference, for DLA~B1228--113 this estimate is consistent with $[\rm{M/H}]=-0.03^{+0.18}_{-0.13}$ estimated from the absorption-based mass-metallicity relation \citep{Moller2013}.
The latter is not applicable for DLA~J0918+1636, where the more likely DLA galaxy is at a closer projected separation to the quasar \citep{Fynbo2013}.} 
While the dispersion around the fundamental relation is very small (0.05~dex) at low redshifts, the dispersion increases to 0.2--0.3 dex at $z \gtrsim 2$ \citep{Mannucci2010}. Taking this dispersion into account, the estimated host galaxy metallicities are consistent with a solar value. This supports the choice of an $\alpha({\rm CO})$ conversion factor of $4.36\; {\rm M_{\odot}(K \; km \; s^{-1} \; pc^2)^{-1}}$. However, if the galaxies have a subsolar metallicity, $\alpha({\rm CO})$ could be greater than the assumed value.

Based on the above, we estimate $\delta_{\rm GDR} = 167^{+298}_{-94}$ for DLA~B1228--113 and $136^{+243}_{-77}$ for DLA~J0918+1636, where the errors include a 0.2~dex uncertainty on the metallicity and a 0.15~dex scatter on the metallicity-$\delta_{\rm GDR}$ relation.

The dust mass is related to the total gas mass by the following relation \citep{Magdis2012a}:
\begin{equation}
\delta_{\rm GDR} \ M_{\rm dust} = M_{\rm gas} \equiv M_{\rm H\textsc{i}} + M_{\rm mol}
\end{equation}

Recent H{\sc i} 21-cm line stacking experiments in star-forming galaxies at $z \approx 1$ have found that, on average, $M_{\rm H\textsc{i}} \approx M_{\rm mol}$ \citep{Chowdhury2020, Chowdhury2021}. This yields
\begin{equation}
\delta_{\rm GDR} \ M_{\rm dust} \approx 2 \times M_{\rm mol}
\end{equation}

Using the dust masses from our MBB fits, we obtain molecular gas masses reported in Table~\ref{tab:MolGasMass}. 
The main uncertainties of this method include the assumption that the galaxies follow the Fundamental Mass Metallicity relation mentioned, the fraction of molecular gas in the total gas mass, and the dust mass absorption coefficient.
Moreover, $\delta_{\rm GDR}$ is calibrated to local main-sequence galaxies. 
\citet{Magdis2012a} argue that assuming no redshift evolution in the dust-to-gas mass ratio is unlikely to be a major concern for their method. They also find that high redshift main-sequence galaxies and SMGs follow approximately the same relation although SMGs tend to have a slightly higher $\delta_{\rm GDR}$ at a given metallicity.

%For DLA~PKS~B1228--113 the molecular gas mass derived from the dust mass is lower than the previously reported molecular gas masses from mid-$J$ CO transitions but consistent within the uncertainties.
For both galaxies we find that  $M_{\rm mol}^{\rm dust}$ is significantly lower than $M_{\rm mol}^{\rm CO}$ if we assume $r_{31} = 0.55$ \citep{Kanekar2020}. 
However, applying the updated $r_{31} = 1$ line luminosity ratio we find that the molecular gas masses derived from the two methods agree within the errors for DLA~B1228--113. In the case of DLA~J0918+1636, $M_{\rm mol}^{\rm dust}$ is lower than $M_{\rm mol}^{\rm CO}$ even for $r_{31} \approx 1$, although formally consistent within the errors. For this galaxy, the thermalised CO excitation conditions up to the $J = 4$ level suggest that $\alpha_{\rm CO}$ may be lower than $4.36  \ {\rm M_\odot}$~(K~km~s$^{-1}$~pc$^2$)$^{-1}$, since $\alpha_{\rm CO}$ is proportional to $\sqrt{n}/T$. If so, this would imply a lower $M_{\rm mol}^{\rm CO}$, in even better agreement with the $M_{\rm mol}^{\rm dust}$ value.

%\textbf{We note that given the thermalised excitation conditions up to $J = 4$ in DLA~J0918+1636, the $\alpha_{\rm CO}$ conversion factor could be closer to $\approx 1\; {\rm M_{\odot}(K \; km \; s^{-1} \; pc^2)^{-1}}$ as seen for example in (U)LIRGS \citep{Papadopoulos2012}.}
%

The assumptions discussed above, their validity, and the consequences for the final estimates have been the focus of dedicated in-depth investigations over the years \citep[][and many others]{Magdis2012a, Scoville2016}. Given the current data availability, more sophisticated calculations are beyond the scope of this work.

\section{Conclusions}

We study the CO spectral line energy distributions (CO SLEDs) of absorption-selected galaxies previously detected in a low $J$ rotational CO transition \citep{Fynbo2018, Neeleman2018, Kanekar2018, Kanekar2020}. The absorbers were originally selected to have high metallicity, and therefore are likely to trace high stellar mass host galaxies \citep{Christensen2014}.
We detect all but one targeted higher $J$ rotational transitions in the observed galaxies. 
Using an automated emission line finder we do not detect any additional galaxies associated with the absorbers in our observations. 
The CO line ratios in two out of four absorption-selected galaxies are low and consistent with the CO line ratios observed in the Milky Way inner disk. 

Two absorption-selected galaxies show elevated CO line ratios with respect to the other absorption-selected galaxies in this sample. 
The lowest redshift galaxy in our sample (DLA~J2335+1501 at $z_{\rm abs} = 0.6798$) shows a CO SLED similar to galaxies on and above the star-forming main sequence at higher redshift, but also consistent with that of the Milky Way. DLA~B0551--366 ($z_{\rm abs} = 1.9622$) too has a CO SLED consistent with that of the Milky Way inner disk, and below the CO SLED of main sequence galaxies at $z \sim 1.5$.

The other two galaxies, DLA~B1228--113 at $z = 2.1929$ and DLA~J0918+1636 at $z=2.5832$, show CO line ratios above that of the Milky Way inner disk. The former shows excitation conditions below the average of main-sequence galaxies at similar redshifts, while the latter shows thermalised excitation in the CO(4--3) transition which we identify as the peak of the CO SLED.
 
Our results show that the excitation conditions in the ISM of absorption-selected galaxies can range from low excitation similar to the inner disk of the Milky Way up to thermal excitation of the mid-$J$ levels. 
Combined with the results from \citet{Klitsch2019}, who presented the first study of CO SLEDs in absorption-selected galaxies at $z\sim0.5$, we now have a sample of seven absorption-selected galaxies with information on their CO SLED. Four of these are at intermediate redshifts $z \sim 0.5 - 0.7$ and three at high redshift, $z \gtrsim 2$). At intermediate redshifts, the range of ISM excitation conditions reflect a range of galaxy types [one potential starburst, two galaxies hosting an active galactic nucleus out of which one is potentially a starburst galaxy \citep{Klitsch2019} and one main-sequence galaxy]. At $z \sim 2$, two of the galaxies presented here have measured stellar masses and SFRs, and lie within the spread of the star-forming main-sequence at their redshifts \citep{Whitaker2014}.
Absorption-selected galaxies at $z \sim 2$ show a range of excitation conditions comparable to that found in main-sequence galaxies at similar redshifts \citep{Brisbin2019, Boogaard2020, Valentino2020a, Henriquez-Brocal2021}.

%\NK{I don't think the next sentence makes sense, because the previous sentence is that the absorption-selected galaxies show a range of excitation conditions similar to that found in emission-selected galaxies. I would urge dropping the first sentence of the next paragraph; it's not needed.}

%Perhaps the range of different excitation conditions inferred for the galaxies are a consequence of the different selection methods. 
Absorption-selected galaxies are detected by a combination of the neutral gas cross-section as well as the luminosity of the absorber host galaxy which again is correlated with the galaxy- and absorber metallicity. 
Several studies have compared the properties of absorption- and emission-selected samples. 
\citet{Krogager2017} find that an absorption-selected galaxy sample covers a larger span of the underlying galaxy luminosity function compared to flux-limited samples. 
\citet{Rhodin2018} find that absorption-selected galaxies at $z \sim 0.7$ follow the main-sequence at that redshift. 
However, at the high-mass end [log($M_{\star}/\text{M}_{\odot}) > 10.5$], absorption-selected galaxies tend to fall below the star-forming main-sequence. 
At $z\sim2$ \citet{Rhodin2021} report that the absorption-selected galaxies identified to date belong to the category of main-sequence galaxies at that redshift, albeit extrapolated to a lower-mass range ($10^8-10^{10}$ M$_{\odot}$) compared to emission-selected samples. 
Finally, based on the molecular gas mass derivations and modest SFRs, \citet{Kanekar2018} report a low star formation efficiency in absorption-selected galaxies at $z \sim 0.7$ compared to emission-selected main-sequence galaxies at $z = 0$ and $z = 2$.

Overall, absorption-selected galaxies have been found to belong to a range of different galaxy types with stellar masses towards the lower mass end compared to emission-selections. 
We conclude that the range of excitation conditions mirror the spread in galaxy types. 
Whether a specific galaxy type is overrepresented as suggested by \citet{Klitsch2019} can only be addressed with a statistically significant sample.
A remaining question is also how the variety of excitation conditions and CO SLEDs in absorption-selected galaxies are connected to other physical properties of the galaxies, and whether there are systematic variations of the CO SLEDs with redshifts.

For DLA~B1228--113 and DLA~J0918+1636, we 
 investigated whether the $L'_{\rm CO (1-0)}$ and the molecular gas mass derived from different scaling relations to the dust luminosity (this work) and mid-$J$ CO transitions \citep{Kanekar2020} are consistent with each other.
%We present a derivation of  $L'_{\rm CO}$ and the molecular gas mass based on empirical relations with the dust emission fitted using a modified black body for two of the galaxies towards PKS~B1228--113 and J0918+1636. 
For both galaxies, we find that  $L'_{\rm CO(1-0)}$ derived from the dust luminosity is
%consistent for DLA~PKS~B1228--113. 
%For DLA~J0918+1636, $L'_{\rm CO(1-0)}$ derived from the dust luminosity is 
lower than $L'_{\rm CO(1-0)}$ derived from the mid-$J$ CO transitions by a factor of $\sim 2$ when assuming a sub-thermal line luminosity ratio of $r_{31} = 0.55$. 
However, applying a line luminosity ratio of $r_{31} = 1$, we find that $L'_{\rm CO(1-0)}$ derived from the two different methods are in reasonable agreement. The molecular gas masses derived from the dust scaling and CO emission also agree within the uncertainties of the respective conversions when assuming $r_{31} = 1$ for DLA~B1228--113 and for DLA~J0918+1636. It thus appears likely that the molecular gas masses of these two absorption-selected galaxies are a factor of $\approx 2$ lower than earlier estimates in the literature, based on sub-thermal excitation of the mid-$J$ rotational levels.
%We note that the molecular gas mass $M_{\rm mol}^{\rm CO}$ of DLA~J0918+1636 could be lower and closer to $M_{\rm mol}^{\rm dust}$ if we apply a conversion factor of $\alpha_{\rm CO} \approx 1\; {\rm M_{\odot}(K \; km \; s^{-1} \; pc^2)^{-1}}$.

It is important to emphasize that our study is  limited by the coverage of only a few observed CO transitions for each galaxy, and especially does not include the J(1--0) transition. %, and especially of the low $J$ rotational transitions. 
As the CO line excitation is a non-linear, non-monotonic function, there may be ambiguities in interpreting the measured line excitation ratios when these are anchored at transitions above the CO(1--0) transition, as we have done here.

%\NK{I would advise dropping the next paragraph. It's not adding anything to the paper and is suddenly talking about emission-selected samples.}

%We clearly require better studies of CO SLEDs in normal star-forming galaxies as well as starburst galaxies at a range of redshifts including a larger sample of galaxies and more CO transitions to significantly improve the molecular gas mass measurements based on higher $J$ CO rotational transitions. 

\section*{Acknowledgements}

A.K.~gratefully acknowledges support from the Independent Research Fund Denmark via grant number DFF 8021-00130.
F.V.~acknowledges support from the Carlsberg Foundation Research Grant CF18-0388 ``Galaxies: Rise and Death''. 
N.K.~acknowledges support from the Department of Atomic Energy, under project 12-R\&D-TFR-5.02-0700.
J.P.U.F.~acknowledges support from the Carlsberg Foundation.
The  Cosmic  Dawn  Center  (DAWN)  is  funded  by  the Danish National Research Foundation under grant No.140.
M.N.~acknowledges support from ERC Advanced grant 740246 (Cosmic\texttt{\char`_}Gas). 
%\textbf{Any more acknowledgements?}
This paper makes use of the following ALMA data: ADS/JAO.ALMA\#2019.1.01457.S. ALMA is a partnership of ESO (representing its member states), NSF (USA) and NINS (Japan), together with NRC (Canada), MOST and ASIAA (Taiwan), and KASI (Republic of Korea), in cooperation with the Republic of Chile. 
The Joint ALMA Observatory is operated by ESO, AUI/NRAO and NAOJ. 
This research made use of Astropy\footnote{\url{http://www.astropy.org}} a community-developed core Python package for Astronomy\citep{astropy2013, astropy2018}.
This research has made use of NASA's Astrophysics Data System.

\section*{Data Availability}

The data underlying this article are available in the respective observatories online archives. 
Used ALMA data is available through the ALMA Science Archive (\url{https://almascience.eso.org/asax/}).
%VLT/MUSE data is available through the ESO SciencenArchive Facility (\url{http://archive.eso.org/cms.html}) 
%and HST data through the Hubble Legacy Archive (\url{https://hla.stsci.edu}).
Project IDs are given in Section~\ref{Section2}.
Based on observations collected at the European Southern Observatory under ESO programme 197.A-0384.



\bibliographystyle{mnras}
\bibliography{MyLibraryNoLink} % if your bibtex file is called example.bib

%\begin{thebibliography}{99}
%\end{thebibliography}

%%%%%%%%%%%%%%%%%%%%%%%%%%%%%%%%%%%%%%%%%%%%%%%%%%

%%%%%%%%%%%%%%%%% APPENDICES %%%%%%%%%%%%%%%%%%%%%

\appendix

\section{Continuum images and SED fits.}

\begin{figure}
\includegraphics[width = 0.49\linewidth]{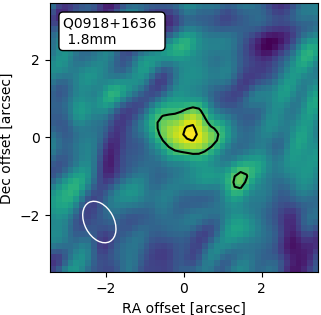}
\includegraphics[width = 0.49\linewidth]{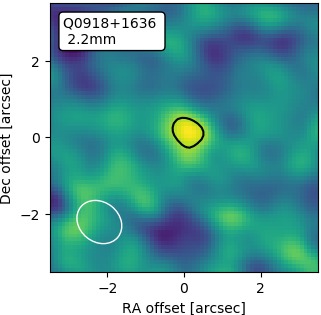}
\includegraphics[width = 0.49\linewidth]{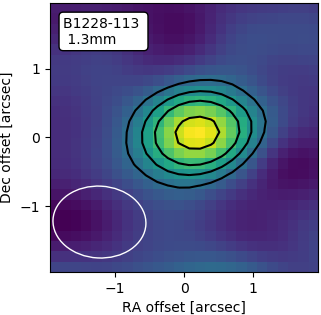}
\caption{ALMA continuum images of DLA~B1228-113 and DLA~J0918+1636. The central wavelength of the continuum image is indicated at the top left of each panel. The first contour in each panel is at FILL$\sigma$, with successive contours increasing by FILL. \label{FigContImage}. }
\end{figure}

This appendix provides information on the SED fits to the optical, near-IR and sub-mm continuum data obtained with {\tt MAGPHYS} \citep{daCunha2008}.

\begin{figure}
\centering
\includegraphics[width = 1.\linewidth]{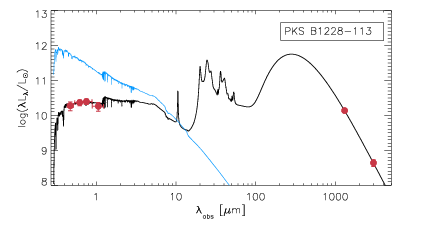}
\includegraphics[width = 1.\linewidth]{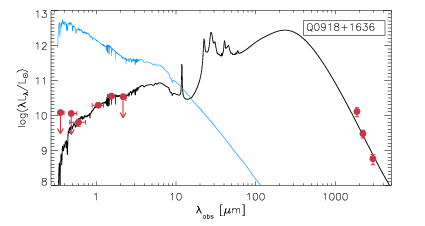}
\caption{SED fits obtained with {\tt MAGPHYS} for DLA~B1228-113 and DLA~J0918+1636.
\label{Fig:appendixfig}}
\end{figure}

In Fig.~\ref{FigContImage} we show the sub-mm continuum images of DLA~B1228--113 and DLA~J0918+1636. The optical and near-IR data for DLA~J0918+1636 were presented originally in \citet{Fynbo2018}; these values are included in Table~\ref{tab:appendix_phot}, for completeness. For DLA~B1228--113, optical magnitudes are derived from MUSE data extracted from the spectral data cube within an aperture of 2 arcsec. The coadded 1-dimensional spectrum is multiplied with the SDSS filter transmission curves to derive AB magnitudes. The near-IR photometry is presented in \citet{Neeleman2018}. The optical and near-IR photometric data were corrected for Galactic reddening \cite{Schlafly2011} and, along with the ALMA continuum flux densities, were used as inputs to {\tt MAGPHYS}. Fig.~\ref{Fig:appendixfig} present the best fit SED models for the two galaxies, while Table~\ref{tab:magphys_out} the resulting best fit values of the stellar mass, the dust mass, and the SFR, along with the 68\% confidence intervals.

\begin{table}
\centering
\caption{Photometry of DLA~B1228--113 and DLA~J0918+1636}
\begin{tabular}{lcc}
\hline
Band   &   J0918+1636 &   B1228--113  \\
\hline
  $u$    &  $>$26.5 (3$\sigma$) & --- \\
  $g$    &  $>$26.2 (3$\sigma$) & 25.3$\pm$0.3 \\ 
  $r$    &  ---                 & 24.7$\pm$0.2 \\
  $i$    &  ---                 & 24.4$\pm$0.2 \\
  $F$606W  &  26.59$\pm$0.25    & ---\\
  $F$105W  &  24.70$\pm$0.13    & 24.36$\pm$0.3 \\
  $F$160W  &  23.63$\pm$0.07    & --- \\
  $K_S$  &  $>$23.3 (3$\sigma$) & --- \\ 
  \hline
   ALMA 1.33 mm  &  ---          & $627\pm50$ \\
   ALMA 1.86 mm  &  565$\pm$162  & --- \\
   ALMA 2.21 mm  &  156$\pm$39   & --- \\
   ALMA 2.95 mm  &  40$\pm$13    &  46$\pm$10 \\
     \hline
\end{tabular}
\flushleft
\textit{Notes:} For the optical and near-IR data of J0918+1636, a 2 arcsec aperture diameter was used. All optical and near-IR magnitudes are in AB units \citep{Fynbo2018}. The table values have not been corrected for Galactic reddening. Continuum flux densities in the ALMA bands are in units of $\mu$Jy.
\label{tab:appendix_phot}
\end{table}

\begin{table}
\centering
\caption{Best-fit values of parameters from {\tt MAGPHYS} obtained when fixing the redshift to that of the two galaxies. Uncertainties represent 68\% confidence intervals.}
\label{tab:magphys_out}
\begin{tabular}{lcc}
\hline
    &   J0918+1636 &   B1228--113  \\
\hline
 SFR [$M_\odot$/yr]  & $229^{+151}_{-91}$      & $87^{+39}_{-28}$ \\
 log ($M_*/M_\odot$) &  $10.49^{+0.16}_{-0.11}$ & $10.13^{+0.34}_{-0.23}$ \\
 log ($M_{\mathrm{dust}}/M_{\odot}$) & $8.62^{+0.16}_{-0.15}$ &  $8.74^{+0.10}_{-0.12}$ \\
     \hline
\end{tabular}
\end{table}


%%%%%%%%%%%%%%%%%%%%%%%%%%%%%%%%%%%%%%%%%%%%%%%%%%

% Don't change these lines
\bsp	% typesetting comment
\label{lastpage}
\end{document}